\documentclass[showpacs,preprintnumbers,amsmath,amssymb, nofootinbib, prd]{revtex4}
\usepackage{amsmath,amssymb,graphicx}
\usepackage{epstopdf}
\usepackage {amssymb}
\newcommand{\nc}{\newcommand}
\nc{\ba}{\begin{eqnarray}}
\nc{\ea}{\end{eqnarray}}
\newcommand\be{\begin{equation}}
\newcommand\ee{\end{equation}}

\newcommand\mPl{{M_{\rm Pl}}}

\newcommand{\md}{\mathrm{d}}

\nc{\x}{{\bf{x}}}
\newcommand{\bmk}{\mathbf{k}}
\newcommand{\bmx}{\mathbf{x}}

\begin{document}

\title{Cosmological Perturbations in Inflationary Models with \\ Anisotropic Space-Time Scaling in Lifshitz Background}

\author{Mohsen Alishahiha$^{1}$}
\email{alishah-AT-mail.ipm.ir}
\author{Hassan Firouzjahi$^{2}$}
\email{firouz-AT-mail.ipm.ir}
\author{Kazuya Koyama$^{3}$}
\email{kazuya.koyama-AT-port.ac.uk}
\author{Mohammad Hossein Namjoo$^{1}$}
\email{mh.namjoo-AT-mail.ipm.ir}
\affiliation{$^1$School of Physics, Institute for Research in 
Fundamental Sciences (IPM),
P. O. Box 19395-5531,
Tehran, Iran}
\affiliation{$^2$ School of Astronomy, Institute for Research in 
Fundamental Sciences (IPM),
P. O. Box 19395-5531,
Tehran, Iran}
\affiliation{$^3$ Institute of Cosmology \& Gravitation, University of Portsmouth, Portsmouth, PO1 3FX, UK}

\date{\today}

\begin{abstract}
\vspace{0.3cm}

Models of inflation in a gravitational background with an anisotropic space-time scaling are studied. The background is a higher-dimensional Lifshitz throat with the anisotropy scaling 
$z\neq 1$. After the dimensional reduction, 
the four-dimensional general covariance is explicitly broken to a three-dimensional spatial diffeomorphism. As a result the cosmological perturbation theory in this set up with less symmetries have to be formulated.  We present the consistent cosmological perturbation 
theory for this set up.  We find that the effective four-dimensional gravitational wave perturbations propagate with a  different  speed than the higher dimensional gravitational 
excitations. Depending on the model parameters, for an observer inside the throat, 
the four-dimensional gravitational wave propagation can be superluminal. We also find that  the Bardeen potential and the Newtonian potential are different. This can have interesting observational consequences for lensing and  CMB fluctuations.  Furthermore, we show that at the linearized level the inflaton field excitations vanish.

\vspace{0.3cm}

\end{abstract}

\maketitle

\section{Introduction}

Inflation has emerged as the leading theory for early universe and structure formation \cite{Guth:1980zm} which is strongly supported by recent Planck observations 
\cite{Ade:2013zuv, Ade:2013uln}.  Simple models of inflation predict almost scale-invariant, almost Gaussian and almost adiabatic perturbations inn cosmic microwave background (CMB) which are very well consistent with cosmological observations.
However, despite these observational successes, inflationary paradigm is at the phenomenological level and there is no deep theoretical understanding of mechanism behind inflation and the nature of inflaton field.

There have been many works to embed inflation in the context of high energy physics such as string theory, for a review see  \cite{HenryTye:2006uv,Cline:2006hu,Burgess:2007pz,McAllister:2007bg,Baumann:2009ni, Mazumdar:2010sa}. In particular, brane inflation is an interesting model of inflation in string theory \cite{dvali-tye, Alexander:2001ks, collection, Dvali:2001fw}. In these scenarios a mobile brane is moving inside the string compactification in the presence of background branes, anti-branes and fluxes. In some of these scenarios,  the interaction between the mobile branes and the background anti-branes is the driving source for inflation \cite{dvali-tye,Alexander:2001ks,collection,Dvali:2001fw, Kachru:2003sx, Firouzjahi:2003zy, Burgess:2004kv, Buchel, Iizuka:2004ct, Firouzjahi:2005dh}. 
Inflation in this picture ends when the brane and anti-brane annihilate each  other resulting a copious productions of cosmic (super)strings which can be detected observationally \cite{Sarangi:2002yt, Majumdar:2002hy, Copeland:2003bj, Firouzjahi:2006vp, Jackson:2004zg}. DBI inflation \cite{Silverstein:2003hf, Alishahiha:2004eh} is another realization of inflation from string theory in which the mobile brane moves ultra-relativistically inside a warped throat \cite{Chen:2004gc, Shandera:2006ax, Bean:2007eh, Bean:2007hc, Thomas:2007sj, Huston:2008ku, Cai:2008if}. A non-trivial predictions of DBI inflation is generating large equilateral type  non-Gaussianities \cite{Chen:2006nt, Seery:2005wm, Chen:2010xka}
which can be detected observationally \cite{Ade:2013ydc}. 

In \cite{Alishahiha:2011yh} an extension of DBI inflation in a Lifshitz throat is studied. The background is a five-dimensional theory in which the time coordinate and the space coordinates scale differently under extra dimension throat coordinate $r$. In \cite{Alishahiha:2011yh} the inflation at the homogenous and isotropic FRW background is studied. Furthermore, it is shown that the four-dimensional general covariance is explicitly broken to a 
three-dimensional spatial diffeomorphism. Therefore, the cosmological perturbation theory in
this setup with less symmetries has to be revisited which is the aim of this work.  

The rest of the paper is organized as follows. In Section \ref{setup}
we present our setup and in Section  \ref{Einstein} the Einstein equations are presented for this background. In Section \ref{tensor} the tensor excitations are studied followed by the scalar excitations in Section \ref{Scalar-perts}. In section \ref{power} we obtain the curvature perturbation and the gravitational anisotropy power spectrum followed by discussion in 
section \ref{discussions}. Some technical issues are relegated to appendices.

\section{The Setup}
\label{setup}

Here we present our setup and  briefly review the results in \cite{Alishahiha:2011yh}.


Our background consists a Lifshitz throat in a string theory compactification. The Lifshitz geometry has attracted considerable attentions recently
in the context of non-relativistic AdS/CFT correspondence where it may 
provide a gravity description for Lifshitz fixed point. We note that Lifshitz fixed
points appear when we are dealing with a physical system at critical point
with anisotropic scale invariance in which the space and time scale differently 
 \cite{Hertz:1976zz}
\be\label{Lif}
t\rightarrow \lambda^zt,\;\;\;\;\;\;x_i\rightarrow \lambda x_i \, ,
\ee
in which $t$ and  $x_i$ respectively are the  time and space coordinates. 
The corresponding critical points are known as Lifshitz fixed points.

It is natural to look for  gravity duals of Lifshitz fixed points in the light of AdS/CFT correspondence \cite{Maldacena:1997re}.   The gravity descriptions of  
Lifshitz fixed points have been studied in \cite{Kachru:2008yh} in which
the metric  invariant under the scaling \eqref{Lif} is obtained to be
\be\label{BacLif}
d s^2=-\left(\dfrac{r}{L} \right)^{2z} \md t^2+ \left(\dfrac{r}{L} \right)^2 \md \bmx^2+\left(\dfrac{L}{r} \right)^2 \md r^2 \, ,
\ee
where $L$ is  the curvature radius of the the ``Lifshitz throat''
and $r$ is the extra dimension  radial coordinate. One can easily check that the metric (\ref{BacLif}) is invariant under the scaling \eqref{Lif} and subject to 
$r \rightarrow \lambda^{-1} r$.

It should be stressed that the ansatz (\ref{BacLif})  may not be a trivial solution of the five-dimensional Einstein equation. In principle there are other fields, such as a massive gauge field, which should be added into the action in order to support the Lifshitz geometry (\ref{BacLif}). 
The corresponding geometry  may also be obtained from a pure gravitational theory by adding higher derivative terms to the Einstein-Hilbert action\cite{AyonBeato:2010tm}. 
Also, unlike the case of AdS throat which may be easily constructed in a string theory setup using brane construction, it is not clear how to  construct a Lifshitz throat 
in string theory. We note, however, that  a string theory realization of Lifshitz background has been constructed in \cite{Hartnoll:2009ns}  in the context of strange metallic holography. For more studies
see also \cite{Balasubramanian:2010uk, Donos:2010tu, Gregory:2010gx, Cassani:2011sv}.
As a result one can not consider our inflationary setup as a top-down approach. Here we shall follow the phenomenological approach and assume that in principle our setup with a Lifshitz throat  can be constructed in string theory. 

In our picture the Lifshitz throat is extended in a localized region of string compactification 
and it is smoothly glued to the bulk of Calabi-Yau  (CY) compactification which is Lorentz invariant as usual. The picture is similar to warped compactifications considered in many phenomenological models such as in  \cite{Kachru:2003sx}. The difference is that the throat, now instead of being AdS, is a Lifshitz background with the anisotropic scaling $z>1$. 
In this view, the inflaton field is a mobile brane which moves ultra-relativistically inside the Lifshitz throat. 
As usual, one may imagine that inflation ends when the mobile brane is annihilated with a background anti-brane. The Lifshitz throat is extended in the region $r_0 < r < R$ in which 
$r_0$ indicates the infra-red (IR) cutoff of the throat while $R$ is the UV cutoff of the throat. 
It is assumed that at $r=R$, the Lifshitz throat is smoothly glued to the bulk of CY compactification which is Lorentz invariant as in conventional models.


Having presented our setup, we promote the Lifshitz metric \eqref{BacLif} into a cosmological background. The background FRW metric is given by
\ba
\label{FRW-metric}
ds^2 = -\left(\dfrac{r}{L} \right)^{2z} \md t^2+  a(t)^2 \left(\dfrac{r}{L} \right)^2 
\md x^i \md x^j+  \left(\dfrac{L}{r} \right)^2 \md r^2
\ea
After compactification over the internal volume ${\cal V}$, one obtains the standard 4D FRW metric \cite{Alishahiha:2011yh}
\ba
ds^2 = - dt^2 + a(t)^2 d \bmx^2 \, .
\ea
In our analysis below, we start from the 5D metric ansatz such as in Eq. (\ref{FRW-metric}) 
with its perturbations included later on and obtain the corresponding Einstein equations. 

As in standard DBI inflation, the D3-brane action is given by DBI and Chern-Simons terms \cite{Alishahiha:2011yh}\footnote{ We note that in lack of the brane construction of  the Lifshitz geometry, the Chern-Simons contribution to the following action should be considered
as a phenomenological insight.} 
\ba 
\label{action}
S=-T_3 \int \md^4 x  \left(\dfrac{r_b}{L} \right)^{3+z} \left(\sqrt{1-\left(\dfrac{L}{r_b}\right)^{2+2z}\dot{r_b}^2}-1 \right)  \, ,
\ea
in which $T_3$ is the D3-brane tension and $r_b(t)$ is the position of brane inside the Lifshitz throat.  The canonically normalized field is given by
\ba
\label{dif-eq}
\dot{\phi} \equiv \sqrt{T_3} \, \dot{r_b}  \, \left(\dfrac{r_b}{L} \right)^{\frac{1-z}{2}}  . 
\ea
After adding the potential term $V(\phi)$ the matter action is given by \cite{Alishahiha:2011yh}
\ba 
\label{DBI-Lif}
S=-\int \md^4 x \, a(t)^3 \left[  f^{-1} \left(\sqrt{1-f \dot{\phi}^2} -1\right) + V(\phi) \right]  \, ,
\ea 
where
\ba
\label{f-def}
f(\phi)=\begin{cases}
T_3^{-1} \left(\frac{\mu_z}{\phi} \right)^\alpha &        \text{if}  \, \, z \neq 3
\\
\\
T_3^{-1} e^{-\frac{6 \phi}{\mu_3}} & \text{if}  \, \, z=3
\end{cases}
\ea
with
 $\alpha \equiv \dfrac{2(3+z)}{3-z}  \, .$
Moreover the parameters $\mu_z$ and $\mu_3$ are defined via
\ba
\label{muz3}
\mu_3 \equiv \sqrt{T_3} L \quad \quad   , \quad \quad
\mu_z \equiv \dfrac{2 \mu_3}{3-z}   = \dfrac{2\sqrt{T_3}}{3-z}\, L  \quad (z\neq 3) \, .
\ea

The background expansion equations are obtained to be 
\ba
\label{back-FRW}
3 M_P^2 H^2  = \rho \quad , \quad
\dot \rho + 3 H (\rho + p) =0 \, ,
\ea
in which $H= \frac{\dot a}{a}$ is the Hubble expansion rate, $\rho$ and $p$ respectively are the energy density and the pressure
\ba
\label{p}
\rho = f^{-1} (\gamma -1)+ V
\quad , \quad 
p=f^{-1}( 1- \gamma^{-1}) -V \,  ,
\ea
and $\gamma$ is the Lorentz factor defined by
\ba
\label{Lorentz}
\gamma \equiv \dfrac{1}{\sqrt{1-f \dot{\phi}^2}}  \, .
\ea
The effective four-dimensional Planck mass, $M_P$, is given by \cite{Alishahiha:2011yh}
\ba
\label{MP}
M_P^2 \equiv \kappa_5^{-2} \int_{\cal V} d\, r \, \left(\frac{r}{L}\right)^{2-z} \, ,
\ea
in which $\kappa_5$ is the 5D gravitational coupling and 
${\cal V}$ is the volume of the compactification.  One can obtain sufficient e-foldings of inflation with appropriate form of $V(\phi)$ subject to slow-roll conditions \cite{Alishahiha:2011yh}.

Also from the brane action (\ref{DBI-Lif}) one can obtain  the Klein-Gordon (KG) equation for the inflaton field 
\ba
\label{KG-back}
 \ddot{\phi}+3H\gamma^{-2}\dot{\phi}+\gamma^{-3} \left( V'+ \dfrac{f'}{2 f^2} (1-\gamma)^2 (\gamma+2) \right)=0 \, ,
\ea 
where the prime denotes the derivative with respect to scalar field $\phi$. One can also check that the KG equation is not independent of Einstein equations (\ref{back-FRW}).

Following \cite{Shandera:2006ax}, one can cast the background equations (\ref{KG-back}) and
(\ref{back-FRW}) into Hamilton-Jacobi forms which are more suitable for analytical purposes. Since $\phi$ is monotonically decreasing 
as time goes by, we can use $\phi$ as the clock and express the physical parameters in terms of
$\phi$. This yields 
\ba
\label{HJ}
3 M_P^2 H(\phi)^2 &=& V(\phi)+f^{-1}(\gamma(\phi) -1) \nonumber \\ 
\gamma(\phi) &=& \sqrt{1+4 M_P^4 f(\phi) H^{'}(\phi)^2}  \nonumber \\ 
\dot{\phi}(\phi) &=& \dfrac{-2 M_P^2 H'}{\gamma(\phi)}  
\ea 
where $H'= \partial H/\partial \phi$. The system of Eq.  (\ref{HJ}) can be solved in the speed limit in which $\gamma \gg 1$, for the details see \cite{Alishahiha:2011yh}. 

In our analysis so far, we have not specified the form of the potential $V$ driving inflation. As argued in \cite{Alishahiha:2011yh}, phenomenologically  we may assume 
$V(r) \sim r^n$ with an arbitrary value of parameter $n$. This form of potential may originate from the back-reactions of the mobile branes with the background fluxes and the volume modulus.  Note that the case $z=1, n=2$ corresponds to conventional model of DBI inflation
with the potential $m^2 \phi^2/2$ which is vastly studied in the literature.  In  Section \ref{power}  we consider the cosmological predictions of our model for different values of $z$ and $n$.

\section{4D Einstein Equations}
\label{Einstein}

Having specified our background inflationary set up, we are ready to consider cosmological perturbation in this set up. However, to study the perturbation equations we need the general form of the Einstein and Klein-Gordon equations in this background. Since our background is not 4D Lorentz-invariant, we have to obtain these equations independently of the known results in standard cosmology literature.

Here we obtain the  general effective 4D Einstein equations. The full five-dimensional action includes both the gravity and matter sectors 
\ba 
S=S_G+S_M
\ea 
The gravitational part is a trivial Einstein-Hilbert action in 5D. 
\ba
S_G=\frac{1}{2\kappa_5^2}\int d^5 x \sqrt{-G} \, \, ^{(5)}R  
\ea
where $\kappa_5^{-1}$ is the 5D gravitational mass scale and $^{(5)}R$ is the 
5D Einstein-Hilbert term.  As for the matter sector, we will not be specific here. As an example, the Lifshitz geometry can be obtained from massive gauge fields \cite{Taylor:2008tg}. 
All that is required is that the matter sector supports the Lifshitz geometry in 5D as given in Eq. (\ref{BacLif}).

Motivated by the Fefferman-Graham like coordinates for the Lifshitz geometry\cite{Taylor:2008tg},  our ansatz for 
the metric perturbations, consistent with the Lifshitz symmetry,  is \cite{Alishahiha:2011yh}
\ba
\label{Lif-cosmo}
d s^2=g_{00}\left(\dfrac{r}{L} \right)^{2z} \md t^2+ g_{ij}\left(\dfrac{r}{L} \right)^2 
\md x^i \md x^j+ 2 g_{0i} \left(\dfrac{r}{L} \right)^{z+1} \md t\,  \md x^i +
\left(\dfrac{L}{r} \right)^2 \md r^2
\ea			
where $g_{\alpha \beta}$ are functions of $x^\mu$. Note that in this notation 
$G_{MN}$ is the 5D metric, while $g_{\alpha \beta}$ is the 4D metric. The capital letters $M, N, ...$ represent the 5D coordinate indices while the greek symbols $\mu, \nu, ...$ indicate the 4D coordinate indices. For a general metric ansatz with an arbitrary scaling for 
$\md t\,  \md x^i $ part of the metric see Appendix \ref{general-metric}.

One way to obtain the 4D fields equations is to perform the dimensional reduction and calculate the effective action for gravity. This program was performed in \cite{Alishahiha:2011yh} in the gauge where $g_{0i}=0$. However, as we shall see later on, $g_{0i}$ are independent dynamical variables and can not be gauged away. This is because the Lifshitz solution 
(\ref{BacLif}) is not invariant under boost and it is invariant only under the three-dimensional rotation. That is, the symmetry transformation of our background (\ref{Lif-cosmo}) is
\ba
\label{3D-diff}
t \rightarrow \tilde t(t) \quad , \quad
\x \rightarrow \tilde \x ( \x) \, .
\ea
As a result the action obtained in \cite{Alishahiha:2011yh}, although correct for terms containing $g_{00} $ and $g_{ij}$, should be supplemented with the additional terms containing $g_{0i}$ terms. However, as we shall see in next sections, the 
condition $g_{0i}=0$ is the only consistent solution in our setup and therefore the results in \cite{Alishahiha:2011yh} for dimensionally reduced gravitational action are valid at the end. 

For non-zero $g_{0i}$ it is a difficult task to calculate the gravitational action to second order in terms of $g_{\alpha \beta}$. Instead we use the variational method which is more straightforward. In the variational method, we vary the 5D gravitational action in terms of
4D metric $g_{\alpha \beta }$ as given in Eq. (\ref{Lif-cosmo}) 
\ba
\delta G_{MN} = \frac{\delta G_{MN}}{\delta g_{\alpha \beta}} \delta g_{\alpha \beta} \, .
\ea 
For this purpose, the following formulas are helpful
\ba
\label{detG}
\sqrt{-G} = \left(\dfrac{r}{L} \right)^{z+2} \sqrt{-g}
\ea
and
\ba
\label{G-inverse}
G^{00} = \left(\dfrac{r}{L} \right)^{-2z} g^{00}\quad , \quad
G^{0i} = \left(\dfrac{r}{L} \right)^{-(1+z)} g^{0i} \quad , \quad
G^{ij} = \left(\dfrac{r}{L} \right)^{-2} g^{ij}
\ea
Equipped with these formulas and varying the action in terms of $g^{\alpha \beta}$
one can check that the 4D Einstein equations are
\ba
\label{Gupdn}
\kappa_5^{-2} \int_{\cal V} dr \left(\dfrac{r}{L} \right)^{(z+2)} {\cal{G}}^0_0&=&T^0_0
 \nonumber\\
\kappa_5^{-2} \int_{\cal V} dr \left(\dfrac{r}{L} \right)^{z+2} {\cal{G}}^i_j&=&T^i_j \nonumber\\
\kappa_5^{-2} \int_{\cal V} dr \left(\dfrac{r}{L} \right)^{(2z+1)} {\cal{G}}^0_i&=&T^0_i \, .
\ea
Here ${\cal{G}}^{\mu}_{ \nu}$ is the 5D Einstein tensor and the integration is over the compactification volume ${\cal V}$. Furthermore, 
$T_{\mu \nu}$ is the symmetric energy-momentum tensor defined via the 4D metric $g_{\alpha \beta}$
\ba
\delta S_{M} = -\frac{1}{2} \int d^4 x \sqrt{-\bar g} \delta g^{\alpha \beta} T_{\alpha \beta} \, .
\ea
Note the different scalings with $z$ 
in Eq. (\ref{Gupdn}) which multiply the components of ${\cal{G}} ^\mu _\nu$
and will have crucial implications in our analysis below.
The set of equations in Eq. (\ref{Gupdn})  is our starting point to calculate the perturbation analysis.  For this purpose we should specify the matter action to obtain $T_{\mu \nu}$.
We also note that there is the $rr$ component of Einstein equation in 5D but since we do not vary the $rr$ component  of metric, we do not obtain additional constraint from the $rr$ component of Einstein equation. Procedure similar to this logic, yielding Eq. (\ref{Gupdn}),
was also employed in \cite{Csaki:1999mp} in the context of Randall-Sundrum cosmology.

As mentioned above the brane action is not the full part of matter sector action.
There are other background fields to support Lifshitz solution. In our treatment below, we do
not perturb these background 5D fields. We assume that these background fields contribute to effective four-dimensional cosmological constant term which vanishes at the background level. As we shall see this  treatment of 5D  matter fields reproduces the standard results in the limit where $z=1$ and is expected to be the case when $z> 1$.

The action for the mobile brane moving in geometry  (\ref{Lif-cosmo})  is \cite{Alishahiha:2011yh}
\ba
S_{(b)}=\int d^4 x  \sqrt{-g} \, {\cal{L}}_{(b)}  \, ,
\ea
in which
\ba
\label{dbibrane}
{\cal{L}}_{(b)}  =-f^{-1} \left[ \left( 1+f \, g^{00} \dot \phi^2 
 +h g^{ij} \partial_i \phi \partial_j \phi +2 {\it{l}} g^{0i}  \dot \phi \partial_i \phi  \right)^{1/2}-1\right]-V \ \nonumber\\
\ea
with  $f$ defined as in Eq. (\ref{f-def}),  $\ell(\phi) = \sqrt{f(\phi) h(\phi)} $
and 

\ba
\label{h-def}
h(\phi)=\begin{cases}
T_3^{-1} \left(\frac{\mu_z}{\phi} \right)^{\alpha'} &        \text{if}  \, \, z \neq 3
\\
\\
T_3^{-1} e^{-\frac{2 \phi}{\mu_3}} & \text{if}  \, \, z=3
\end{cases}
\ea
with 
 \ba
  \alpha' \equiv \dfrac{2(5-z)}{3-z} \, .
\ea

Varying the brane action in Eq.  \eqref{dbibrane} one can read the energy momentum tensor
\ba
\label{T-eq}
T_{\alpha \beta} = g_{\alpha \beta} {\cal L} + \Gamma \left[ \dot \phi^2 \delta^0_\beta  \delta^0_\alpha
+ \delta^i_\alpha \delta^j_\beta F(\phi)^2 \partial_i \phi \partial_j \phi   + 
\left( \delta^0_\alpha \delta^i_\beta + \delta^0_\beta \delta^i_\alpha  \right) F(\phi) \dot \phi 
\partial_i \phi \right] \, .
\ea
in which 
\ba
\label{F-def}
F(\phi) \equiv \sqrt{ \frac{f(\phi)}{h(\phi)} } = \left( \frac{r_b}{L}\right)^{z-1} \, .
\ea
and
\ba
\Gamma\equiv  \left( 1+f \, g^{00} \dot \phi^2 
 +h g^{ij} \partial_i \phi \partial_j \phi +2 {\it{l}} g^{0i}  \dot \phi \partial_i \phi  \right)^{-1/2}  \, .
\ea
Note that in the isotropic limit with $z=1$, we have $F=1$. As we shall see below, the fact that $F \neq 1$ plays an important role in our analysis. Also the function $\Gamma$ is defined such that it reduces to the Lorentz factor $\gamma$ in Eq. (\ref{Lorentz})
at the background FRW.  One can easily check that the Einstein equations  (\ref{Gupdn}) with 
the energy momentum tensor given in Eq. (\ref{T-eq}) reproduce the expected background FRW equations (\ref{back-FRW}) with $M_P$ given in Eq. (\ref{MP}). 

Having obtained the general Einstein equation in Eqs. (\ref{Gupdn}) with $T_{\alpha \beta}$
given in Eq. (\ref{T-eq}), we study the scalar and tensor perturbations in this inflationary background. As we shall see, the scalar perturbations analysis is non-trivial. So we start with the tensor perturbation analysis which proved to be simpler. 

\section{Tensor Perturbations}
\label{tensor}

Here we consider the tensor perturbations where
the metric excitations are given by
\ba
\label{Lif-tensor}
ds^2 = -\left(\dfrac{r}{L} \right)^{2z} \md t^2+ \left(\dfrac{r}{L} \right)^2 
\left( \delta_{ij} + h_{ij} \right) \md x^i \md x^j+  \left(\dfrac{L}{r} \right)^2 \md r^2 \, ,
\ea
where we impose the transverse and traceless gauges on  $h_{ij}$ 
\ba
h^i_i = \partial_i h^i_j =0 \, .
\ea
Here the spatial indices are raised and lowered by $\delta_{ij}$.

One can specifically check that $\delta T^{\mu}_\nu={\delta \cal G}^0_0 = {\delta \cal G}^0_i =0$ and 
\ba
{\delta \cal G}^i_j = \frac{1}{2} \left(\frac{r}{L}\right)^{-2 z} \left( \ddot h_{ij} + 3 H \dot h_{ij} \right) - \frac{1}{2} a^{-2}\left(\frac{r}{L}\right)^{-2 }
 \nabla^2 h_{ij}  \, ,
\ea
in which $\nabla^2 = \partial_i \partial_i$.
Plugging this into (\ref{Gupdn}) yields
\ba
\ddot h_{ij} + 3 H \dot h_{ij} - (1+\beta_0) a^{-2}\nabla^2 h_{ij} =0 \, ,
\ea
where the dimensionless parameter $\beta_0$ is defined via (note that in \cite{Alishahiha:2011yh}, $\beta_0$ is denoted by $\bar \Omega$ )
\ba
\label{beta0}
\beta_0  \equiv -1 + \frac{\int_{\cal V} dr \left(\frac{r}{L}\right)^z}{\int_{\cal V} dr\,  \left(\frac{r}{L}\right)^{2-z}} \, .
\ea
Note that in the limit where $z=1$,  we see that $\beta_0=0$, so $\beta_0$ may be thought as  a measure of 4D Lorentz violation.  As we shall see in the following analysis, $\beta_0$ is a key parameter of our model.

Defining $\hat h_{ij} \equiv a h_{ij}$ and going to the Fourier space we find
\ba
\label{tensor-eq}
\hat h_{ij}'' + \left( c_g^2 k^2 -\frac{2}{\tau^2} \right) \hat h_{ij} =0\, ,
\ea
where the prime denotes derivative with respect to the conformal time $  d\tau = dt/a(t)$
and the gravitational wave (GW) speed $c_g$ is given by
\ba
\label{cs}
c_g^2 \equiv 1+\beta_0 \, .
\ea
Eq. (\ref{cs}) is very interesting. It indicates that the GW propagates with a non-trivial speed $c_g$. In the limit where $z=1$, we have the usual result that $c_g=1$. However, with arbitrary value $z$, $c_g$ can be very different than unity. Note that in our convention, we have set the gravitational wave speed equal to unity in 5D. Therefore, Eq. (\ref{cs}) indicates that in 4D the gravitational waves propagate with a different speed 
than in 5D.  Depending on the value of $z$ and the size of Lifshitz throat, $c_g $ can be even bigger than unity! The value of 
$\beta_0$ is calculated in \cite{Alishahiha:2011yh} which results in
\ba
\label{cs-value}
c_g^2 \simeq   \left\{
\begin{array}{c}
\frac{3-z}{z+1} \left( \frac{R}{L} \right)^{2 (z-1)}  \quad , \quad  ~~~~~~ z<3
\\\\
\frac{z-3}{z+1} \left( \frac{R}{L} \right)^{2 (z-1)} \left(\frac{r_0}{R}
\right)^{z-3}  \quad , \quad  z>3
\\\\
\frac{\left(\frac{R}{L}\right)^4}{4 \left(\ln \frac{R}{r_0} \right)} \quad , \quad  ~~~~~~~~~~~~~z=3
\end{array}
\right. \hspace{0.5cm}
\ea
Here $R$ is the UV cutoff of the Lifshitz throat where it is smoothly glued to the bulk of CY compactification and $r_0$ is the IR cut off of the throat which plays the role of IR TeV brane
in Randall-Sundrum picture \cite{Randall:1999ee}. We also expect that $R \gtrsim L$  while $R/r_0$ to be exponentially large in the light of string flux compactification \cite{Giddings:2001yu}. 

As mentioned above we have set the speed of light and GW propagation in five dimension to unity.
So if $c_g$ is bigger than unity, it indicates the superluminal GW propagation compared
to a 5D observer. Now, we have to see what the 4D speed of light is. For this purpose, consider a 4D observer located at an arbitrary fixed position $r=r_O$ in the Lifshitz throat. Finding the speed of light for this observer
from the condition $ds^2=0$ yields
\ba
\label{c-gamma}
c_{\gamma}^2 = \left(\frac{r_O}{L}\right)^{2 (z-1)}  = F(\phi_O)^2
\ea 
in which $F(\phi)$ is defined in Eq. (\ref{F-def}).  Comparing the 4D GW speed $c_g$ given in Eq. (\ref{cs-value}) and the 4D photon speed $c_\gamma$ in Eq. (\ref{c-gamma}) we see that if $r_O$ is sufficiently near the IR end of the throat then $c_g >c_\gamma $ and an observer located at $r=r_O$  experiences a superluminal GW propagation in four-dimensional sense.
Similar ideas was proposed in 
the context of Randall-Sundrum brane world cosmology \cite{Csaki:2000dm}.

\section{Scalar Perturbations}
\label{Scalar-perts}

In this section we consider the scalar metric perturbations following the conventions of \cite{Bassett:2005xm}.
In terms of $g_{\alpha \beta}$
given in Eq. (\ref{Lif-cosmo}) the scalar perturbations are denoted by
\ba
g_{00} = - (1+ 2 A) \quad , \quad 
g_{0i} = a(t) \partial_i B  \quad , \quad
g_{ij}= a(t)^2 \left[ (1- 2 \psi) \delta_{ij}
+ 2 \partial_i \partial_j E \right] \, .
\ea
In total we have four scalar metric degrees of freedom, $A, B, \psi$ and $E$. However,  the theory enjoys only the three-dimensional spacial diffeomorphism as given in Eq. (\ref{3D-diff}). One can easily check that all four variables $A, B, \psi$ and $\dot E$ are gauge invariant under the three-dimensional spatial diffeomorphism in Eq. (\ref{3D-diff}). On top of this, the inflaton field excitations $\delta \phi$ is also gauge invariant under Eq. (\ref{3D-diff}).
As a result, we should expect to have five independent equations for the five physical perturbations $A, B, \psi, E$ and $\delta \phi$. This  is in contrast to usual situation in standard cosmological perturbation theory in which due to gauge freedom one ends up with two scalar metric degrees of freedom plus $\delta \phi$.

The components of perturbed 5D Einstein tensor, $\delta {\cal G}^M_N$, are given in Appendix \ref{G-MN}.
The $(ij) \,  i\neq j, (0i), (ii) $ and $(00)$ components of perturbed Einstein equations in Eq. (\ref{Gupdn}) are 
\ba
\label{dbi-ij}
(1+ \beta_0) (\psi-A) + H \chi + \dot \chi + a (\dot B + 2 H B) (\beta_1-1) =0&& \\
\label{dbi-0i}
(1-\beta_1)(\dot \psi + H A) + \beta_2 a B 
 - \frac{F(\phi) }{2 M_P^2} \gamma  \dot \phi \delta \phi =0&&\\
\label{dbi-ii}
\ddot \psi +  H ( 3 \dot \psi + \dot A) + (3 H^2 + 2 \dot H ) A - 
\frac{1}{2M_P^2} \left[ \gamma \dot \phi \delta \dot \phi - \gamma \dot \phi^2 A +\left( \dfrac{f'}{2\gamma f^2}(1-\gamma)^2 - V_{, \phi}\right) \delta \phi
\right] =0&& \\
\label{dbi-00}
3H (\dot \psi + H A) - \frac{\vec \nabla^2}{a^2} \left[ ( 1+ \beta_0 ) \psi + 
H a^2 \dot E - a H  (1-\beta_1) B   \right]   \hspace{6cm}
 \nonumber \\
- \frac{1}{2 M_P^2} \left[ \gamma^3 \dot \phi(\dot \phi A - \delta \dot \phi) +\left(-\dfrac{f'}{2 f^2}(1-\gamma)^2(\gamma+2)  - V_{, \phi} \right) \delta \phi \right] =0&&
\ea
in which we have defined 
\ba
\chi \equiv a^2 (\dot E - \frac{B}{a} ) \, .
\ea
Also we have defined the dimensionless parameters $\beta_1$ and $\beta_2$ which comes from various integrations over the compactification volume ${\cal V}$ in Eq. (\ref{Gupdn}) via
\ba
\label{beta12}
\beta_1  \equiv 1 - \frac{\int_{\cal V} dr \left(\frac{r}{L}\right)}{\int_{\cal V} dr\,  \left(\frac{r}{L}\right)^{2-z}} \quad , \quad
\beta_2  \equiv \frac{1-z}{L^2} \frac{\int_{\cal V} dr \left(\frac{r}{L}\right)^{z+2}}{\int_{\cal V} dr\,  \left(\frac{r}{L}\right)^{2-z}} \, .
\ea
Note that in the AdS limit where $z=1$, both $\beta_1$ and $\beta_2$ vanish. One can also check that the set of equations (\ref{dbi-ij})- (\ref{dbi-00}) reduces to the standard equations in conventional perturbation theory \cite{Bassett:2005xm}.

Finally, the Klein Gordon equation is
\ba
\label{dbi-KG}
\delta \ddot  \phi + 3 H  \left(1+s \right) \delta \dot \phi +  {\cal{C}}_1 \delta \phi - \frac{F(\phi)^2}{a^2 \gamma^2} \vec \nabla^2 \delta \phi = {\cal{C}}_2 A + \dot \phi \dot A + 3 \gamma^{-2} \dot \phi  \dot \psi -
 \dot \phi  \, \dfrac{\vec \nabla^2}{\gamma^2}   \left( \dot E - \frac{F(\phi)}{a}  B \right)  \, ,
\ea 
in which
\ba 
{\cal{C}}_1 &&\equiv \left(\dfrac{f'}{2 f^2}\right)' \gamma^{-3} (1-\gamma)^2 (2+\gamma)+\dfrac{3 H f'}{2 f}\dot \phi (s+1-\gamma^{-2})  +\gamma^{-3 }V_{,\phi \phi} 
\\
\nonumber
{\cal{C}}_2  &&\equiv \dfrac{3 f'}{2}\gamma^2 \dot \phi^4  + 3 H \dot \phi (1+\gamma^{-2}) +\ddot \phi (3 \gamma^2-1)  \, ,
\ea 
and we have defined a new slow roll parameter as
\ba
s \equiv \dfrac{\dot \gamma}{H \gamma}  \, .
\ea
Again we note that for the AdS background where $z=1$  and $F(\phi)=1$ we obtain the standard perturbed Klein-Gordon equation. 

In conclusion, we have five independent equations (\ref{dbi-ij}), (\ref{dbi-0i}),  (\ref{dbi-ii}), (\ref{dbi-00}) and (\ref{dbi-KG}) for five variables $A, \psi, B, E $ and $\delta \phi$. In the AdS limit of $z=1$, we restore the standard 4D diffeomorphism invariance so we have only three independent variables, two metric scalar perturbations plus $\delta \phi$. However, when the 
4D diffeomorphism is broken to a subset of 3D rotational invariance, all $A, B, \psi$, $ E$ and $\delta \phi$ become physical degrees of freedom so we have five physical variables in total.

\subsection{Solving the System of Equations}

Here we provide the solutions for the set of five equations (\ref{dbi-ij}), (\ref{dbi-0i}),  (\ref{dbi-ii}), (\ref{dbi-00}) and (\ref{dbi-KG}) for five variables $A, \psi, B, E $ and $\delta \phi$. It looks formidable to find analytical solutions for  this system of coupled equations.
However, things become considerably simple as we shall see below.

Using Eqs. \eqref{dbi-0i} and \eqref{dbi-ii} one obtains the following relation 
\ba
\label{B-phi-eq} 
\beta_2 {(a^4 B)}^. = \frac{1}{2M_P^2}\left(  \left(F+\beta_1 -1 \right) a^3 \gamma \dot \phi \delta \phi \right)^.
\ea
which can be solved easily to yield 
\ba
\label{dbi-B-sol}
a B =  \frac{1}{2M_P^2} \frac{F+\beta_1 -1}{\beta_2} \gamma  \dot \phi \delta \phi  \, .
\ea 
To obtain this equation use was made of the background KG equation as well as the equation 
$ \dot H=- \dfrac{\gamma \,  \dot \phi^2}{2 M_P^2} $. Also a term like $c/a^3$ with $c$ a constant  can be added to either side of Eq. \eqref{dbi-B-sol}. However, treating $B$ and $\delta \phi$ as perturbations which can be turned off at arbitrarily initial time, we conclude that $c=0$. Eq. \eqref{dbi-B-sol} plays a crucial rule in our analysis below. Note that in the isotropic limit in which $z=1$, $\beta_1 = \beta_2=0$
and $F=1$, Eq. \eqref{B-phi-eq} is trivially satisfied.  This is a manifestation of Bianchi identity. However, in our case with $z \neq 1$, the Bianchi identity does not hold, since the left-hand side of Eq. (\ref{Gupdn}) is an integration of the Einstein tensor along the extra dimension.

Using \eqref{dbi-B-sol} to eliminate $B$ in \eqref{dbi-0i} results in
\ba
\label{dbi-A}
 \dot \psi + H A 
 = \frac{1}{2 M_P^2} \gamma  \dot \phi \delta \phi \, .
\ea
Surprisingly this is the same as in standard DBI inflation. 

To simplify further the system, let us define the new variables $\Sigma$ and $\delta{ \phi_\psi}$
as follows
\ba
\label{Sigma-def}
\Sigma &\equiv & (1+\beta_0)\psi +H \chi - a H (1-\beta_1) B
\\
\label{delta-def}
\delta{ \phi_\psi} &\equiv &\delta \phi + \frac{\dot \phi}{H} \psi \, .
\ea 
As we shall see in next sub-section $\delta{ \phi_\psi}$ is related to the
curvature perturbations on comoving surface ${\cal R}$ whereas $\Sigma$ is similar to the Bardeen potential $\Psi$.


Now manipulating the independent Einstein equation \eqref{dbi-ij},   \eqref{dbi-00} and \eqref{dbi-A} one obtains the following set of coupled equations for $\Sigma$ and $\delta \phi_\psi$
\ba
\label{coupled1}
\label{dbi-sys1}
\dot \Sigma + H (1+\epsilon) \Sigma &=& (1+\beta_0) \dfrac{\gamma \dot \phi}{2 M_P^2} \delta \phi_\psi \\
\label{dbi-sys2}
 \frac{\nabla^2}{a^2} \Sigma &=& \frac{\gamma^3 }{2 M_P^2} \left(\dot \phi\,  \dot {\delta \phi_\psi} 
-(\ddot \phi +\epsilon \dot \phi H)\delta \phi_\psi \right) \, .
\ea
Manipulating further Eqs. \eqref{dbi-sys1} and \eqref{dbi-sys2} one obtains the following decoupled system of equations for $\Sigma$ and $\delta \phi_\psi$
\ba
\label{dbi-phi-psi-eq}
\ddot{\delta \phi_\psi} +3H(1+s) \dot{ \delta \phi_\psi} + {\cal{J}} \delta \phi_\psi -\dfrac{1+\beta_0}{\gamma^2}\dfrac{\nabla^2}{a^2} \delta \phi_\psi =0 \\
\label{dbi-Sigma-eq}
\ddot{\Sigma}-2 H^2(2 \epsilon-\eta)\Sigma+H(1+2 \eta -2\epsilon)\dot{\Sigma}- \dfrac{1+\beta_0}{\gamma^2} \dfrac{\nabla^2}{ a^2}\Sigma=0 \,  ,
\ea
in which we used
\ba
\epsilon \equiv - \frac{\dot H}{H^2} \quad , \quad 
\eta \equiv  \epsilon - \dfrac{\ddot H}{2 H \dot H }\quad , \qquad 
\dot \epsilon =  2 H \epsilon (2 \epsilon -\eta)
\ea
and defined 
\ba
{\cal{J}} \equiv  -\dfrac{H}{\gamma^3 a^3 \dot \phi} \, \, \dfrac{d}{dt} \left( \dfrac{\gamma^3 a^3}{H}(\ddot \phi +\epsilon H \dot \phi)\right) 
={\cal{C}}_1 -4 \epsilon \dfrac{H \ddot \phi}{\dot \phi} -H^2 \epsilon(3+4 s +3 \gamma^{-2}+2 \epsilon) \, .
\ea
Eqs. (\ref{dbi-phi-psi-eq}) and (\ref{dbi-Sigma-eq}) are our key equations to find curvature perturbation power spectrum in next sections. 

Looking into $\delta \phi_\psi$ and $\Sigma$ equations in  (\ref{dbi-phi-psi-eq}) and 
(\ref{dbi-Sigma-eq})  we find that the scalar perturbation sound speed, $c_s$,  is 
\ba
\label{cs-eq}
c_s^2 = \frac{1+\beta_0}{\gamma^2} = \frac{c_g^2}{\gamma^2} \, .
\ea
This is an interesting formula. Similar to standard DBI inflation, the scalar perturbations sound speed is suppressed by the factor $1/\gamma$ compared to GW speed. However, note that in conventional theories with 4D Lorentz invariance where $c_g=1$, Eq. (\ref{cs-eq}) is translated into the well-known formula $c_s = 1/\gamma$.

In obtaining the system of Eqs. (\ref{dbi-phi-psi-eq}) and (\ref{dbi-Sigma-eq}) we have used all
four independent Einstein equations \eqref{dbi-ij},   \eqref{dbi-0i}, \eqref{dbi-ii}, and \eqref{dbi-00}. However, so far we have not used the KG equation  (\ref{dbi-KG}). We have to check whether it carries extra information.  Starting with KG-equation \eqref{dbi-KG} and using \eqref{dbi-phi-psi-eq} and after a long but straightforward calculations one obtains the following equation 
\ba
\label{dbi-constraint1}
\nabla^2 \left[ (F^2 -1 -\beta_0) \delta \phi + \dot \phi (F+\beta_1 - 1) a B \right]=0 \, ,
\ea
or by using \eqref{dbi-B-sol} 
\ba 
 \label{dbi-cond2}
   \left( F^2 -1 -\beta_0 + \frac{\epsilon H^2}{\beta_2} \left( F+\beta_1 - 1
\right)^2  \right) \nabla^2 \delta \phi  =0  \, .
\ea
This is a constraint equation for $\delta \phi$. In the isotropic limit in which $z=1$, $\beta_1 = \beta_2=0$ and $F=1$, Eq. (\ref{dbi-cond2}) is trivially satisfied. This is a manifestation of the fact that in standard theories with explicit 4D general covariance the Bianchi identity holds. As a result once the Einstein equations are satisfied, the KG equation is trivially satisfied. However, in our case at hand, the 4D general covariance is explicitly broken to a subset of three dimensional rotational invariance. As a result we have five 
independent equations, four from  Einstein equations and one from  KG equation, for five physical perturbations $A, B, \psi, E$ and $\delta \phi$.

The pre-factor  of $\delta \phi$ in Eq. (\ref{dbi-cond2}) is a dynamical quantity and does not necessarily vanish at the background. As a result, Eq. (\ref{dbi-cond2}) is translated into $\nabla^2 \delta \phi=0$. In a spatially isotropic background with the appropriate boundary conditions at infinity the only consistent solution for this Laplace equation is $\delta \phi=0$. 
Alternatively, in an isotropic background one can expand $\delta \phi $ in Fourier space
$\delta   \phi  = \sum \delta \phi_\bmk (t)  e^{-i. \bmk . \bmx} $ so $\nabla^2 \delta \phi \sim \sum k^2 \delta \phi_\bmk (t)  e^{-i. \bmk . \bmx}$. As a result the only valid
solution of $\nabla^2 \delta \phi=0$ for each mode $\bmk$ consistent with appropriate boundary conditions at infinity is $ \delta \phi_\bmk$.

 This indicates that at the linear perturbation theory, the scalar field excitations 
decouple from the system \footnote{This is a surprising result which requires deeper understanding.  We note, however, that it might be due to the fact that upon reduction 
from five to four dimensions the scalar field $\phi$ gets a mass dimension which 
depends on $z$.  Indeed the $z$-dependence of the mass dimension for $z\geq 2$ makes the field to be irrelevant perturbation as the mobile brane moves toward the  Lifshitz
throat. Therefore one might suspect  that the dynamics of the scalar field may not be 
important, at least as far as the  linear perturbation theory is concerned. }.
Note also that, from Eq. (\ref{dbi-B-sol}) and $\delta \phi=0$ one  concludes that $B=0$. This also means that at the linear perturbation theory the off-diagonal metric perturbations 
$g_{0i}=  a(t)\partial_i B$ are not excited. As we promised, this validates the analysis in \cite{Alishahiha:2011yh} in which to obtain the effective 4D gravitational action the gauge $g_{0i}=0$ have been used.  In Appendix \ref{app-action} we have provided further insight
on the decoupling of $\delta \phi$ at the level of quadratic action.

It is also interesting to examine the numerical factor inside the bracket in Eq. (\ref{dbi-cond2})
and see how close to zero it can be.  From the definition of $\beta_2$ in Eq. (\ref{beta12}) we have $\epsilon H^2/\beta_2 \sim \epsilon\,  (H L)^2$. In our effective 4D theory the size of extra dimension is much smaller than the 4D length scale so $L \ll 1/H$ or $HL \ll1$ and therefore the term containing $\epsilon $  in Eq. (\ref{dbi-cond2}) is negligible.  As a result 
we are left with the condition $   \left(F(\phi)^2 - 1-\beta_0 \right)  \nabla^2 \delta \phi \sim 0$. On the other hand, $1+\beta_0 =  c_{g}^2$ where we have defined $c_{g}$ as the gravitational wave speed.  Furthermore, the photon speed for the observer localized on the moving brane from Eq. (\ref{c-gamma}) is 
$c_\gamma^2(\phi) = F(\phi)^2$. Therefore, the condition (\ref{dbi-cond2}) is approximately written as $ ( c_{g}^2 - c_{\gamma}^2(\phi) \, ) \nabla^2  \delta \phi \sim 0 $. 
Therefore, in our system where  $c_{g} \neq c_{\gamma}(\phi)$, we conclude that
$\delta \phi=0$.

Now the last equation to be solved is  an equation for $A$ which can be obtained from either of equations  \eqref{dbi-A} or \eqref{dbi-ij}. For example, using Eq. \eqref{dbi-A} with $\delta \phi=0$
we obtain 
\ba
\label{A-sol}
A= -\frac{1}{H} \left( \frac{H}{\dot \phi} \delta \phi_\psi
\right)^. \, .
\ea

In conclusion, our systems of equations are $B= \delta \phi=0$ along with 
Eqs. (\ref{dbi-phi-psi-eq}), (\ref{dbi-Sigma-eq}) and (\ref{A-sol}).
Once we solve $\delta \phi_\psi$
and $\Sigma$, from Eqs. (\ref{Sigma-def}) and (\ref{delta-def})
we can find $\psi$ and $\chi$  and from Eq. (\ref{A-sol}) we find the value of $A$.
However, from now on, we trade $\psi$ and $\chi$ in terms of $\delta \phi_\psi$ and $\Sigma$ which are more physical. 

\subsection{Gravitational Anisotropy}
\label{anisotropy}

As  in standard cosmological perturbation theory, it is useful to work with the comoving curvature perturbation ${\cal R}$ which in the convention of \cite{Bassett:2005xm}
is related to $\delta \phi_\psi $ via
\ba
\label{R}
{\cal R} \equiv \frac{H}{\dot \phi} \delta \phi_\psi = \psi \, ,
\ea 
where the last equation holds because in our model $\delta \phi=0$.
On the other hand, our $\Sigma$ is similar to the Bardeen potential $\Psi$ in the standard 
situation $\Psi \equiv \psi + H \chi - a H B$. One can check that
\ba
\label{Sigma-R}
\Sigma = \Psi + \beta_0 \psi + \beta_1 a H B = \Psi + \beta_0 {\cal R} \, ,
\ea
where the last equation is obtained in our model knowing that $B=0$. Note that in the limit
where $\beta_0 = \beta_1=0$, we have $\Sigma = \Psi$.

Now we prove the important result that for super-horizon modes ${\cal R}$ is conserved. 
With the definition of ${\cal{R}}$ in  Eq.\eqref{R} and using Eq. \eqref{dbi-phi-psi-eq} one can check that in the Fourier space 
\ba
\dfrac{1}{\epsilon \gamma^2  a^3} \dfrac{d}{dt} \left( \dot {\cal{R}_\bmk} \epsilon \gamma^2 a^3 \right)+ \, \dfrac{c_s^2 \, k^2}{a^2}  {\cal{R}_\bmk}=0 \, .
\ea
This means that on super-horizon scales in which $c_s k/a H\rightarrow 0$, the curvature perturbation ${\cal{R}_\bmk}$ is constant outside the sound horizon crossing, where now 
the sound speed is given in Eq. (\ref{cs-eq}).

One novel aspect of our system with Lorentz violation is the appearance of primordial gravitational anisotropy controlled by the difference $\Phi -\Psi$ in which   the gauge invariant Newtonian potential $\Phi$ is defined by
$ \Phi \equiv  - \partial_t \left(\chi - a B \right) = A - \dot \chi $.  In standard theories with explicit 4D general covariance and in the absence of spatial anisotropies the relation $\Phi =\Psi$ holds. As a result, it is a good choice in our model to define the anisotropy field 
\ba
\label{sigma-def}
\sigma \equiv \Phi -\Psi \, .
\ea
Using Eq. (\ref{dbi-ij})  and noting that $B=0$ in our system, one obtains
\ba
\label{sig}
\sigma = \beta_0 (\psi - A) = \beta_0 \left( {\cal R} + \frac{\dot {\cal R}}{H} \right) \, .
\ea
Interestingly enough, we see that the anisotropy field $\sigma$ is controlled by the parameter $\beta_0$.  As we have seen above, on the super-horizon scales $ \cal R$ is conserved and
$\sigma \simeq \beta_0 {\cal R}$ which also from Eq. (\ref{Sigma-R})
leads to $\sigma \simeq \Sigma - \Psi$.
As expected, in standard situation where $\beta_0=0$ we obtain $\sigma=0$.

Observationally, gravitational lensing and Integrated Sachs-Wolfe (ISW) effect are sensitive to
$\sigma$ while galaxy peculiar velocity measurements are determined by the Newtonian potential $\Phi$  \cite{Jain:2007yk, Afshordi:2008rd}. It would be interesting to look for observational implications of our model with  anisotropy field $\sigma \neq 0$.


\begin{figure}
\includegraphics[scale=.9]{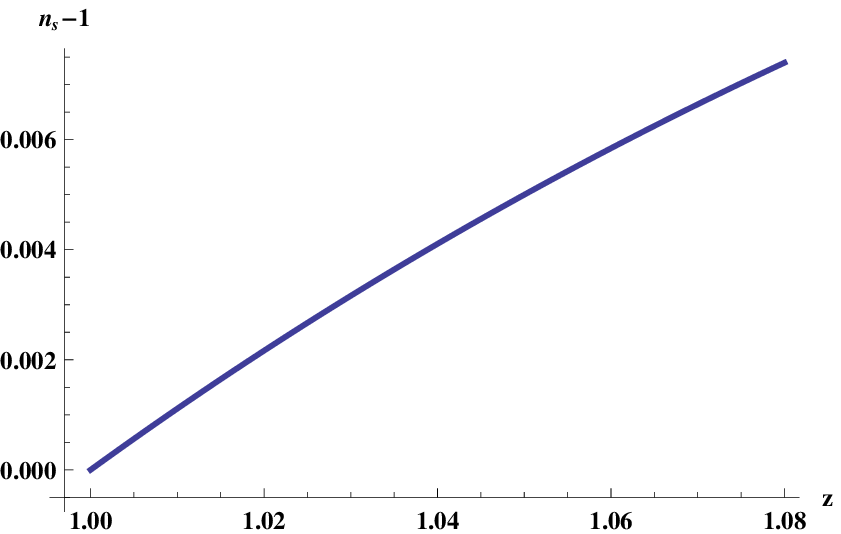}\hspace{0.5cm}
\includegraphics[scale=.9]{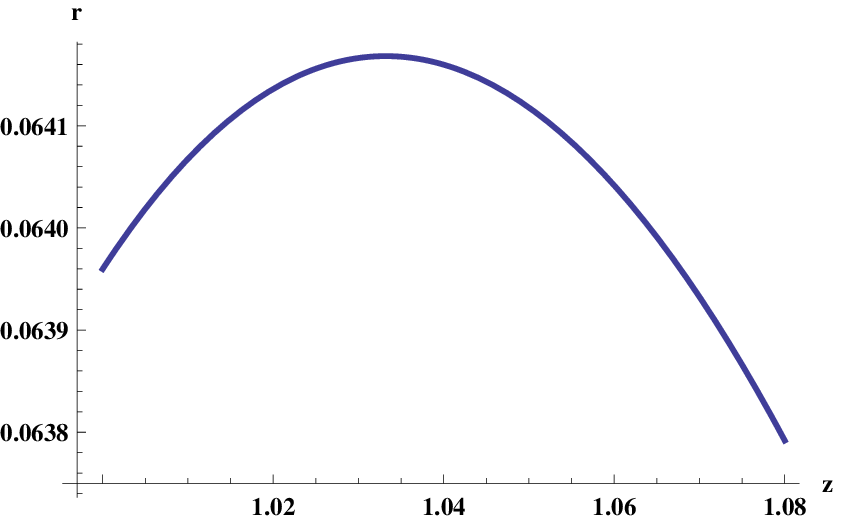} \hspace{0.5cm}
\caption{Here we plot $n_s -1$ and  $r$, the ratio of the tensor to scalar power spectrum, with $n=2$ for the mode
which leaves the sound horizon at $N=5$ e-folds. While $z$ is varying all other parameters and the initial conditions are held fixed. 
The spectral index increases towards blue as $z$ increases while $r$ shows more  non-trivial behavior. }  \vspace{0.7cm}
\label{n2z-b} 
\end{figure}

\section{Power Spectra}
\label{power}

In this section we calculate the power spectrum of the curvature perturbation ${\cal P}_{\cal R}$ and the anisotropy field  ${\cal P}_{\cal \sigma}$.

To solve \eqref{dbi-phi-psi-eq} let us define the Sasaki-Mukhanov variable 
\ba 
v=(\frac{a}{a_0})^{(1+ 3s/2)}\delta \phi_\psi \, ,
\ea
where $a_0$ is the scale factor at some initial time. Going to conformal time  
Eq. \eqref{dbi-phi-psi-eq} reduces to
\ba
v''_\bmk+\left[ c_s^2k^2- \dfrac{\nu^2-1/4}{\tau^2}\right] v_\bmk =0  \, ,
\ea
in which to first order in slow-roll parameters 
\ba
\nu \simeq \dfrac{3}{2}+3\epsilon - \eta + s 
\ea
and we have used $a H \tau (1-\epsilon)\simeq -1 $ and ${\cal{J}} \simeq -3 H^2 (2 \epsilon - \eta - \dfrac{s}{2} ) $. 

With the Bunch-Davies initial condition the solution, as usual, can be written in terms of Hankel function
\ba
v_k \simeq \frac{\sqrt{- \pi \tau}}{2} e^{i\pi(1+ 2 \nu)/4} H_\nu^{(1)} (- c_s k \tau) \, .
\ea
The curvature power spectrum ${\cal P}_{\cal R}$ is
\ba
\label{R-powr}
\langle  {\cal R}_{k} {\cal R}_{k'}\rangle \equiv (2\pi)^{3} P_{{\cal R}}(k)~\delta^3(k+k')
\quad , \quad 
{\cal P}_{\cal R}\equiv \frac{k^{3}}{2 \pi^{2}}P_{\cal R}(k)\, .
\ea
Solving the above standard equation and using \eqref{R} to relate $\delta \phi_\psi$ to 
${\cal R}$ one finds the power spectrum of curvature perturbation as follows 
\ba 
{\cal{P_R}} \simeq \left(1+\beta_0 \right)^{-3/2} 
\left(\dfrac{H^2}{2 \pi \dot \phi}\right)^2 \, . 
\ea
Also the spectral index defined via $n_s -1 \equiv d \ln {\cal P}_{\cal R} /d \ln k$, is obtained to be 
\ba
n_s-1 \simeq -6 \epsilon +2 \eta +s \, .
\ea
We note that  the form of spectral index is the same as the standard case where $z=1$. The difference is due to the $z$-dependence of slow-roll parameters as well as the definition of sound horizon crossing given in Eq. (\ref{cs-eq}).

On the other hand, the power spectrum of  the tensor perturbations  is given by
\ba
{\cal P}_T = \dfrac{2 H^2}{M_P^2 \pi^2}  (1+\beta_0)^{-3/2} \, .
\ea
As a result, one can obtain the tensor to scalar ratio $r$ by
\ba
r = \dfrac{16 \epsilon}{\gamma} \, ,
\ea
which  is also the same as in the standard case.

\begin{figure}[t]
\includegraphics[scale=.9]{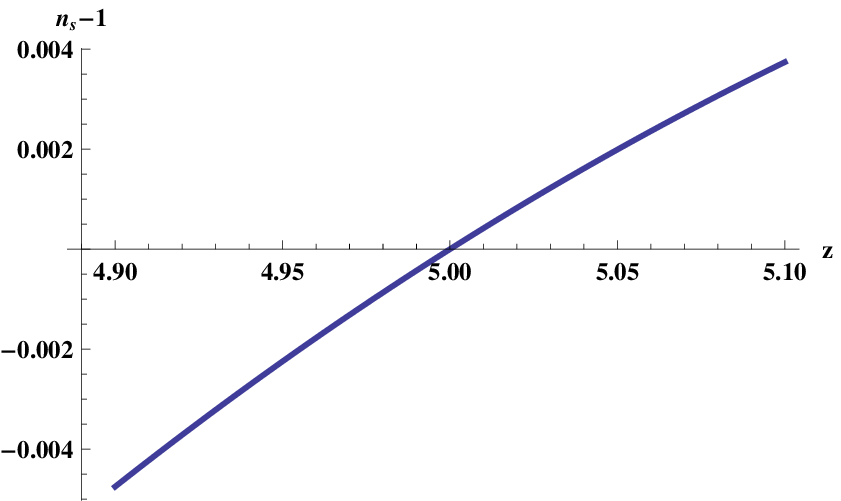} \hspace{1cm}
\includegraphics[scale=.9]{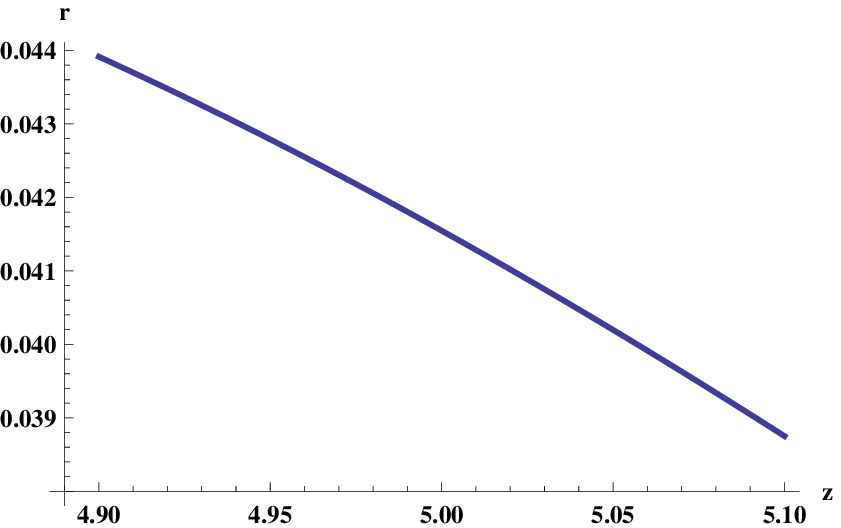}
\caption{The plots of $n_s-1$ and $r$ for the case $n=4$ as function of $z$ for modes which leave the sound horizon at $N=5$ e-folds. As $z$ increases $r$ decreases while,  as expected from Eq. (\ref{ns}), the spectral index is red-tilted, scale invariant and blue-tilted respectively for  $z<5, z=5$ and $z>5$. }
\vspace{0.5cm}
\label{n4z-b} 
\end{figure}

In the speed limit where $\gamma \gg1$, one can find useful expressions for physical parameters. In particular, in the speed limit we have

\ba
\label{ns}
n_s -1 \simeq \frac{2 \epsilon}{n} (3+ z - 2 n) \quad \quad (\gamma \gg 1) \, ,
\ea
in which, as mentioned in Section \ref{setup}, $n$ is the power of inflationary potential in $r$ coordinate
i.e. $V\sim r^n$. In particular, from Eq. (\ref{ns}) we find that for $n=2$, i..e quadratic potential, the spectral index is always blue-tilted. The behaviors  of $n_s$ for $n=2$ and 
$n=4$ are shown in Figures \ref{n2z-b} and  \ref{n4z-b}.

\begin{figure}[t]
\includegraphics[scale=.9]{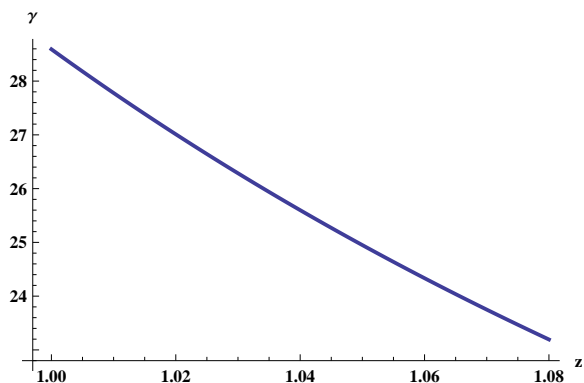}
\caption{  $\gamma$ as a function of $z$ for modes which leave the sound horizon at $N=5$ e-folds for the case $n=2$. The behavior of $\gamma$ for the case of $n=4$ is similar to this plot. }
 \label{n2z-gamma}.  
\end{figure}

The non-Gaussianity parameter $f_{NL}$ in standard DBI inflation is calculated to be \cite{Alishahiha:2004eh, Chen:2006nt} $f_{NL} \sim 0.3 \gamma^{-2}$. We expect that 
the order of magnitude of $f_{NL}$ in our model to be similar to standard DBI case and 
$f_{NL} \sim  \gamma^{-2}$.  To satisfy the Planck constraints on $f_{NL}$ with $c_s \ge 0.07$ \cite{Ade:2013ydc} one concludes that \cite{Alishahiha:2004eh, Chen:2006nt} $\gamma < 14$.  In Figures \ref{n2z-gamma}  the plot of $\gamma$ as a function of $z$ is presented.

\begin{figure}
\includegraphics[scale=.7]{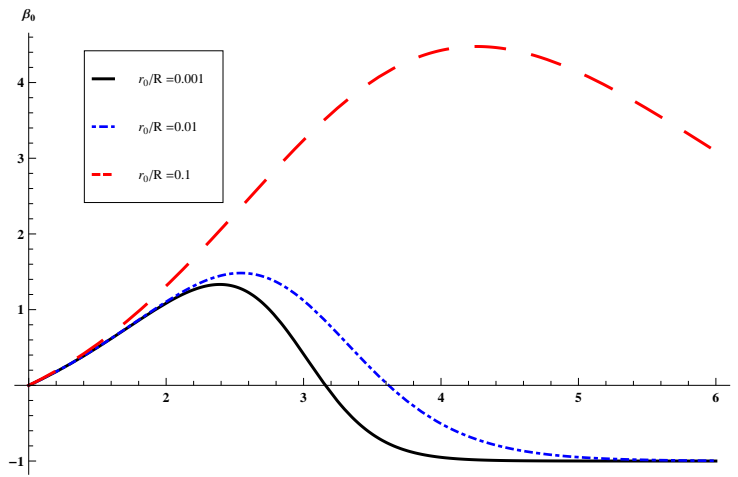} \hspace{1cm}
\caption{ $\beta_0$ as a function of $z$
for different values of the warp factor $r_0/R$. }
\vspace{0.5cm}
\label{beta0-fig} 
\end{figure}

As explained before, one novel aspect of our model is the gravitational anisotropy
in which $\sigma \neq 0$. Now we calculate the anisotropy power spectrum, ${\cal P}_{\sigma}$, on super-horizon scales.  Using Eq. (\ref{sig}) and noting that $\dot {\cal R}=0$ on super-horizon scales yields 
\ba
{\cal P}_{\sigma} = \beta_0^2 {\cal P}_{\cal R} = \frac{\beta_0^2}{(1+ \beta_0)^{3/2}}
\left(\frac{ H^2}{2 \pi \dot \phi}\right)^2 \, .
\ea
Knowing that $c_g^2 = (1+ \beta_0)$, we obtain the following consistency relation between
${\cal P}_{\cal R}$ and ${\cal P}_{\sigma}$
\ba
\label{consistency}
\frac{ {\cal P}_{\sigma}}{{\cal P}_{\cal R}} =  (c_g^2 - 1)^2 = \beta_0^2 \, .
\ea
For example, in standard DBI scenario in which $c_g=c_s$ one obtains 
${\cal P}_{\sigma} =0 $ as expected. The amplitude of ${\cal P}_{\sigma}$ is controlled by
$\beta_0$ or equivalently $c_g$. In Figure \ref{beta0-fig} we plot  $\beta_0$  as a function of $z$.

\section{Discussions}
\label{discussions}

In this work we studied cosmological perturbation in higher dimensional Lifshitz background  with anisotropic scalings in time and space coordinates. The model describes the dynamics of brane inflation in the Lifshitz throat with the anisotropy scaling $z>1$. As we argued, the 4D general covariance is explicitly broken to a subset of three-dimensional rotational invariance. As a result we have four physical metric scalar degrees of freedom in addition to the 
inflaton perturbations $\delta \phi$. After considering the perturbed Einstein and Klein-Gordon equations, we found the unexpected results that at the linear perturbation level $\delta \phi$ as well as $g_{0i}$ excitations are decoupled from the system .

One interesting predictions of our model is the generation of the gravitational anisotropy in which the Bardeen potential and the Newton potential are not equal. Having calculated the power spectrum of the anisotropy field $\sigma = \Phi - \Psi$, we found that it is directly related to the parameter  $\beta_0$ which controls the level of 4D general covariance breaking. 

We have also calculated $n_s$ and $r$ as a function of $z$. Depending on the form of the inflationary potential, $n_s$ can either red- or blue-tilted while $r$ is typically small which is a manifestation of small field range in string theory inflationary model buildings. It is also interesting to calculate non-Gaussianities in this scenario rigorously. However, we expect that 
$f_{NL} \sim \gamma^{-2}$. As a result generating large non-Gaussianities and gravitational anisotropies may be considered two generic features of this scenario.

Another interesting result of our model is that the gravitational wave perturbations and the scalar perturbations  speeds are related via $c_{s} = c_g/\gamma$.  Furthermore, depending on the Standard Model (SM)  observer  position inside the Lifshitz throat, the tensor propagations can be superluminal compared to the photon propagations on the SM brane.  However, the theory is diffeomorphism invariant in 5D so there is no violation of ``causality'' in the five-dimensional sense.

As we argued in Introduction section, we do not have a concrete theoretical realization of this setup in string theory so this work may be considered as a phenomenological exercise for a rigorous string theory background. Having this said, it is interesting that a higher dimension space-time with anisotropy scalings of time and space under the extra dimension coordinate  shows such novel features in effective 4D cosmology.

\section*{Acknowledgement}
We would like to thank D. Baumann, E. Gumrukcuoglu,  J. Lidsey,  S. Mukohyama,
D. Mulryne,  R. Tavakol and D. Tong for useful discussions and  comments. KK is supported by STFC grant ST/H002774/1 and ST/K0090X/1, the European
Research Council and the Leverhulme trust.

\appendix
\section{Components of Einstein Tensor}
\label{G-MN}

Here we present the component of 5D Einstein tensor which will be used in Eq. (\ref{Gupdn}).
One can check that
\ba
\label{G00}
\delta {\cal G}^0_0 &=& 2 \left(\frac{r}{L}\right)^{-2z} \left[ 3 H \dot \psi - H \nabla^2 \dot E+ 3 H^2 A + \frac{H}{a} \left(\frac{r}{L}\right)^{z-1} \nabla^2 B - \frac{1}{a^2}  
\left(\frac{r}{L}\right)^{2(z-1)} \nabla^2 \psi 
\right]\\
\label{G0i}
\delta {\cal G}^0_i &=&-2 \left(\frac{r}{L}\right)^{-2z} \left[  \dot \psi + H A+ 
\frac{(1-z)a}{L^2} \left(\frac{r}{L}\right)^{z+1} B
\right]_{,i} \\
\label{Gii}
\delta {\cal G}^1_1 &=&  -(\partial_y^2 + \partial_z^2) \Omega +2 \left(\frac{r}{L}\right)^{-2z}
\left[ (H^2 + \frac{2 \ddot a}{a}) A +  \ddot \psi + 3 H \dot \psi + H \dot A 
\right] \nonumber\\
\label{G12}
\delta {\cal G}^1_2 &=&\Omega_{, x y}
\ea
where $\Omega$ is defined via
\ba
\Omega \equiv \left(\frac{r}{L}\right)^{-2z} \left[ \ddot E + 3 H \dot E  \right]
- \frac{1}{a} \left(\frac{r}{L}\right)^{-z-1} \left[ \dot B + 2 H B  \right]
+ \frac{1}{a^2} \left(\frac{r}{L}\right)^{-2} (\psi - A)
\ea

\section{A More Generic Metric Perturbations Ansatz}
\label{general-metric}

In obtaining the constraint equation  (\ref{dbi-cond2}) one may
worry whether this conclusion is due to our ansatz in Eq. (\ref{Lif-cosmo}) in which the  Lifshitz scaling symmetry is imposed at the perturbation level.  Here we investigate the general perturbation ansatz which is not restricted to the Lifshitz symmetry at the perturbation level. We find that the constraint equation similar to Eq. (\ref{dbi-cond2}) still holds.  In order to simplify the analysis, we restrict ourselves to slow-roll case where $\gamma =1$. The generalization to the general DBI case will be similar.

A generalization of metric perturbations Eq. (\ref{Lif-cosmo}) without the Lifshitz scaling symmetry is
\ba
\label{Lif-cosmo-g}
d s^2=g_{00}\left(\dfrac{r}{L} \right)^{2z} \md t^2+ g_{ij}\left(\dfrac{r}{L} \right)^2 
\md x^i \md x^j+ 2 g_{0i} \left(\dfrac{r}{L} \right)^{\kappa} \md t\,  \md x^i +
\left(\dfrac{L}{r} \right)^2 \md r^2 \, ,
\ea	
where $\kappa$ is a free parameter. In our analysis in the main text we set $\kappa=z+1$
so the scaling symmetry is preserved at the perturbation level. However, here we consider the arbitrary value of $\kappa$.

With some efforts one can check that the Einstein equations are 
\ba
\label{Gupdn-g}
\kappa_5^{-2} \int_V dr \left(\dfrac{r}{L} \right)^{(z+2)} {\cal{G}}^0_0
\left(1+ \frac{f(r)}{2}  (\nabla B)^2 \right) 
&=&-T^0_0
 \nonumber\\
\kappa_5^{-2} \int_V dr \left(\dfrac{r}{L} \right)^{z+2} {\cal{G}}^i_j
\left(1+ \frac{f(r)}{2}  (\nabla B)^2 \right) &=&-T^i_j  \\
 \kappa_5^{-2} \int_V dr \left(\dfrac{r}{L} \right)^{(z+\kappa)} {\cal{G}}^0_i&=&-T^0_i + \int_V dr  \left(\dfrac{r}{L} \right)^{(z+2)} f(r) \,  {\cal{G}}^0_0 \, g_{0i}
\ea
in which 
\ba
f(r) \equiv 1- r^{2 (\kappa-1-z)} \, .
\ea
In the limit where $\kappa = z+1$, then $f(r)=0$ and we recover Einstein equations as 
in Eq. (\ref{Gupdn}).

Now we present the components of 5D Einstein tensor for arbitrary value of $\kappa $. One can check that there are just some minor modifications and 
\ba
\label{G00-g}
\delta {\cal G}^0_0 &=& 2 \left(\frac{r}{L}\right)^{-2z} \left[ 3 H \dot \psi - H \nabla^2 \dot E+ 3 H^2 A + \frac{H}{a} \left(\frac{r}{L}\right)^{\kappa-2} \nabla^2 B - \frac{1}{a^2}  
\left(\frac{r}{L}\right)^{2(z-1)} \nabla^2 \psi 
\right]\\
\label{G0i-g}
\delta {\cal G}^0_i &=&-2 \left(\frac{r}{L}\right)^{-2z} \left[  \dot \psi + H A+ 
\frac{(2-\kappa) (3+\kappa-z)a}{4 L^2} \left(\frac{r}{L}\right)^{\kappa} B
\right]_{,i} \\
\label{Gii-g}
\delta {\cal G}^1_1 &=&  -(\partial_y^2 + \partial_z^2) \tilde  \Omega +2 \left(\frac{r}{L}\right)^{-2z}
\left[ (H^2 + \frac{2 \ddot a}{a}) A +  \ddot \psi + 3 H \dot \psi + H \dot A 
\right] \\
\label{G12-g}
\delta {\cal G}^1_2 &=&\tilde \Omega_{, x y} \, ,
\ea
in which $\tilde \Omega$ is defined via
\ba
\tilde \Omega \equiv \left(\frac{r}{L}\right)^{-2z} \left[ \ddot E + 3 H \dot E  \right]
- \frac{1}{a} \left(\frac{r}{L}\right)^{\kappa-2z-2} \left[ \dot B + 2 H B  \right]
+ \frac{1}{a^2} \left(\frac{r}{L}\right)^{-2} (\psi - A) \, .
\ea


To calculate $T^{\alpha}_\beta$ we need to calculate the brane  action with the arbitrary value of $\kappa$. We have to calculate $| \bar g_{ab}|$ where $\bar g_{ab}$ is the induced metric on the mobile brane. With some efforts one can show that 
\ba
|\bar g| = r^{2z +6} |g| \left[ 1- f(r) a(t)^{2}\sum_i  (g^{0i})^2  + g^{00} r^{-2z -2} \dot r^2 + r^{-4} g^{ij} \partial_i r \partial_j r + 2 r^{\kappa -2z-4 } g^{0i} \partial_i r \partial_j r
\right] \, .
\ea
In the slow-roll limit, the Lagrangian is
\ba
{ \cal L}  =   -\frac{T_3}{2} r^{z +3}
\left[  g^{00} r^{-2z -2} \dot r^2 +  r^{-4} g^{ij} \partial_i r \partial_j r + 2  r^{\kappa -2 z -4 } g^{0i} \partial_i r \partial_j r - f(r) a(t)^{2} \sum_i ( g^{0i})^2
\right]
\ea
As a result, we have the following action for slowly rolling brane
\ba
 \label{general-metric-g}
S_{(b)}&=&  -\frac{T_3}{2}\int d^4 x \sqrt{-g}  \left(\frac{r_b}{L} \right)^{z+3}
\\ \nonumber
&\times&\left[ g^{00} \left(\frac{r_b}{L}\right)^{-2(z+1)} \dot r_b^2
 +  \left(\frac{r_b}{L}\right)^{-4} g^{ij} \partial_i r_b \partial_j r_b + 2 \left(\frac{r_b}{L}\right)^{(\kappa-4-2z)}  g^{0i} \dot r_b \, \partial_i r_b - f(r) g_{0i} g^{0i} 
 \right] \, .
 \ea
Translating to  the canonically normalized scalar field $\phi$ the action becomes
\ba
\label{brane2-g}
S_{(b)} =\int d^4 x\, \sqrt{-  g } \left[ -\frac{1}{2} g^{00} \dot \phi^2 
 -\frac{F(\phi)^2}{2} g^{ij} \partial_i \phi \partial_j \phi - g^{0i} Q(\phi) \dot \phi \partial_i \phi-\dfrac{T_3}{2} R(\phi)  \sum_i g_{0i} g^{0i}
 -V(\phi)  \right]  \, ,
\ea
where the relation between $\phi$ and $r_b$ is the same as in Eq. (\ref{dif-eq}) and 
\ba
 Q(\phi)= \left(\frac{r_b}{L}\right)^{\kappa-2}
 \qquad 
 R(\phi)= \left(\frac{r_b}{L}\right)^{z+3} f(r_b)   \, .
\ea
Varying the action up to first order in perturbations  we obtain 
\ba
\label{T-stress-g}
T_{\alpha \beta} = g_{\alpha \beta} {\cal L}_{(b)} + \dot \phi^2 \delta^0_\beta  \delta^0_\alpha
+ \delta^i_\alpha \delta^j_\beta F(\phi)^2 \partial_i \phi \partial_j \phi   + 
\left( \delta^0_\alpha \delta^i_\beta + \delta^0_\beta \delta^i_\alpha  \right) 
\left ( Q(\phi) \dot \phi  \partial_i \phi  - T_3 R(\phi) g_{0i}
\right)
\ea
From this we obtain
\ba
T^0_0  &\simeq& \frac{1}{2} g^{00} \dot \phi^2 - V \\
T^i_j & \simeq& \delta^i_j \left( - \frac{1}{2} g^{00} \dot \phi^2 - V \right) \\
T^0_i  & \simeq& - \dot \phi Q \partial_i \phi + T_3 R(\phi) g_{0i} \, .
\ea


Finally, gathering all the above information one can obtain the perturbed Einstein equations.
The $(ij)$ component of Einstein equation for $i \neq j$ in Eq. (\ref{Gupdn-g}) results in 
\ba
\label{eq12-g}
(1+ \beta_0) (\psi-A) + H \chi + \dot \chi + a (\dot B + 2 H B) (\tilde \beta_1-1)=0 \, ,
\ea
where $\chi \equiv a^2 \dot E$.
Note that  $\beta_0$ is the same as in Eq. (\ref{beta0})
while $\tilde \beta_1$ is defined as
\ba
\label{beta01-g}
\tilde \beta_1  \equiv 1 - \frac{\int_{\cal V} dr \left(\frac{r}{L}\right)^{\kappa-z}}{\int_{\cal V} dr\,  \left(\frac{r}{L}\right)^{2-z}} \, .
\ea
From the integrated $(ii)$ component, we find
\ba
\label{eqii-g}
\ddot \psi +  H ( 3 \dot \psi + \dot A) + (3 H^2 + 2 \dot H ) A = 
\frac{1}{2M_P^2} \left[ \dot \phi \delta \dot \phi - \dot \phi^2 A - V_{, \phi} \delta \phi
\right] \, .
\ea
The $(0i)$ and $(00)$ components of Einstein equations yield
\ba
\label{eq0i-g}
(1-\tilde \beta_1)(\dot \psi + H A) +(\tilde \beta_2-\dfrac{3}{2}H^2 \beta_\kappa +\frac{T_3 R}{2 M_P^2}) a B 
 = \frac{Q(\phi) }{2 M_P^2} \dot \phi \delta \phi \, ,
\ea
where $\tilde \beta_2$  and $\beta_\kappa$ are  defined via
\ba
\label{beta2-g}
\tilde \beta_2 \equiv  \frac{(2-\kappa) (3+\kappa-z)}{4 L^2} \frac{\int_{\cal V} dr \left(\frac{r}{L}\right)^{z+2}}{\int_{\cal V} dr\,  \left(\frac{r}{L}\right)^{2-z}}  \quad , \quad 
\beta_\kappa \equiv1-\frac{\int_{\cal V} dr \left(\frac{r}{L}\right)^{2 \kappa-3z}}{\int_{\cal V} dr\,  \left(\frac{r}{L}\right)^{2-z}}  \, .
\ea
Note that in the limit $\kappa=z+1$ the new parameter $\beta_\kappa $ vanishes
while $\tilde \beta_2$ reduces to $\beta_2$ as given in Eq. (\ref{beta12}).

Finally for the (0,0) component one has
\ba
\label{eq00-g}
3H (\dot \psi + H A) - \frac{\vec \nabla^2}{a^2} \left[ ( 1+ \beta_0 ) \psi + 
H a^2 \dot E - a H  (1-\tilde \beta_1) B   \right] 
= \frac{1}{2 M_P^2} \left( \dot \phi^2 A - \dot \phi \delta \dot \phi - V_{, \phi} \delta \phi \right) \, .
\ea
Manipulating Eqs. (\ref{eq0i-g}) and (\ref{eqii-g}) one can show that

\ba
{\left(a^4 B (\tilde \beta_2-\dfrac{3}{2}H^2 \beta_\kappa +\frac{T_3 R}{2 M_P^2}) \right)}^. = \frac{1}{2M_P^2}\left( (Q+\tilde \beta_1 -1) a^3 \dot \phi \delta \phi \right)^.
\ea
which yields 
\ba
\label{B-sol-g}
a B \left(\tilde \beta_2-\dfrac{3}{2}H^2 \beta_\kappa +\frac{T_3 R}{2 M_P^2} \right)=  \frac{1}{2M_P^2} (Q+\tilde \beta_1 -1)  \dot \phi \delta \phi  \, .
\ea
This is similar to Eq. (\ref{dbi-B-sol}) for the case $\kappa = z+1$.  Now, plugging this into Eq. (\ref{eq0i-g})  one obtains
\ba
\label{0i-new-g}
\dot \psi + H A = \frac{1}{2M_P^2} \dot \phi \delta \phi \, .
\ea
This is formally the same as Eq. (\ref{dbi-A}) in the slow-roll limit where $\gamma=1$.

The Klein-Gordon equation has the following form
\ba
\label{eqKG-g}
\delta \ddot  \phi + 3 H  \delta \dot \phi + V_{, \phi \phi} \delta \phi - \frac{F(\phi)^2}{a^2} \vec \nabla^2 \delta \phi = - 2 V_{,\phi} A + \dot \phi \dot A + 3 \dot \phi \dot \psi -
 \dot \phi  \, \vec \nabla^2   \left ( \dot E - \frac{Q(\phi)}{a}  B \right) \, .
\ea

Now that all the equations become similar to the case when $\kappa = z+1$, 
one can check that using the Einstein equations into the KG equation (\ref{eqKG-g}) yields 
\ba
\label{constraint1-g}
\nabla^2 \left[ (F^2 -1 -\beta_0) \delta \phi + \dot \phi (Q+\tilde \beta_1 - 1) a B \right]=0 \, .
\ea
Now using \eqref{B-sol-g}  to express $B$ in terms of $\delta \phi$ one obtains
\ba
\nabla^2 \left[\left(F^2-1-\beta_0- \dfrac{\epsilon H^2(Q+\tilde \beta_1-1)^2}{\tilde \beta_2-3H^2 \beta_\kappa /2+T_3R/2 M_{P}^2} \right) \delta \phi \right] 
\ea
As a result, similar to conclusion from Eq. (\ref{dbi-cond2}), $\delta \phi$ decouples regardless of the value of $\kappa $.  
\section{Decoupling of $\delta \phi$ at the level of action}
\label{app-action}

Here we study the decoupling of  $\delta \phi$ in the quadratic action.  To simplify the analysis, we consider the slow-roll limit corresponding to $\gamma =1$. The extension to DBI case will be similar.

The matter action, in slow-roll limit is the following 
\ba
\label{brane}
S_{M} =\int d^4 x\, \sqrt{-  g } \left[ -\frac{1}{2} g^{00} \dot \phi^2 
 -\frac{F(\phi)^2}{2} g^{ij} \partial_i \phi \partial_j \phi - g^{0i} F(\phi) \dot \phi \partial_i \phi - V(\phi)  \right] \nonumber\\
\ea
while, for the gravity sector we have 
\ba
S_G=\dfrac{1}{2 \kappa_5^2}\int d^5 x \sqrt{-G} \, \, ^{(5)}R  
\ea
However, as mentioned in the draft, we should subtract a non-dynamical cosmological constant term from the above Lagrangian, as an effective result of the extra fields. So we effectively have 
\ba
S_G^{eff}=\dfrac{1}{2 \kappa^2}\int d^5 x \sqrt{-G} \, \, \left[ ^{(5)}R -\Lambda \right]  
\ea
where, from background evolution one can obtain \cite{Alishahiha:2011yh}
\ba
\Lambda = -(12+6 z+2z^2)/L^2  \, .
\ea
Now, we can expand the effective gravitational action as well as the matter sector up to second order. After lots of integration by parts and simplifications, and also integrating out the $r$-coordinate, we end up with
\ba
\nonumber
S_2 = \int d^4 x a^3 \biggl{[} M_P^2 &\bigl{[}&  
 -3 \dot \psi^2 +(\beta_0+1) (\nabla \psi/a)^2 + 2 A (\beta_0+1) \dfrac{\nabla^2}{a^2}\psi -6 H A \dot \psi - (3H^2 + \dot H) A^2 
\\ 
 && +2 (\dot \psi + A H) \nabla^2 \left( \dot E -(1-\beta_1) B/a \right) + \hat \beta_2 (\nabla B)^2
\bigl{]}
\\ \nonumber
 +\dfrac{1}{2} &\delta \dot \phi^2 &-\dfrac{F^2}{2 a^2} (\nabla \delta \phi)^2 -\dfrac{1}{2} V'' \delta \phi^2 + 3 \dot \phi \delta \phi \dot \psi -V' A \delta \phi - \dot \phi \dot{\delta \phi} A -\dot \phi \delta \phi \nabla^2 \left(\dot E -F B/a \right)
\biggl{]}
\ea
where we have defined 
\ba 
\label{beta2}
\hat \beta_2 \equiv  \frac{(1-z)^2}{4 L^2} \frac{\int_{\cal V} dr \left(\frac{r}{L}\right)^{z+2}}{\int_{\cal V} dr\,  \left(\frac{r}{L}\right)^{2-z}} 
\ea
which is slightly different from $\beta_2$ defined in the main text, Eq. (\ref{beta12}). 
Interestingly, apart from the form of $\beta_2$ which are different, the above action gives the consistent equations of motion, Eqs. (\ref{dbi-ij}) - (\ref{dbi-00}). It is also consistent with \cite{Garriga:1997wz} in the case $z=1$. 

Note that the fields  $B$ and $A$ are not dynamical and we can solve the corresponding  constraint and plug the solution back to the action. Furthermore, $E$ appears linearly so using the equation of motion obtained by varying the action with respect to $E$ we find an action for two dynamical fields $\psi$ and $\delta \phi$. After this procedure, let us introduce a new variable 
\ba
 \Omega \equiv \delta \phi + \dfrac{H}{\dot \phi} \psi
\ea
which is actually $\delta \phi_\psi$ in the draft. Using the new field $\Omega$ to eliminate $\psi$ from the action, and after many simplifications, we find the following action for two fields $\Omega$ and $\delta \phi$. 
\ba
S = \int d^4 x \, a^3 \biggl{[}  \dfrac{1}{2} \dot{\Omega}^2 -\dfrac{(\beta_0 +1)}{2 a^2} (\nabla \Omega)^2 
-\dfrac{1}{2} \Omega^2[ V'' - \dfrac{1}{\mPl^2 a^3} \dfrac{d}{dt} ( a^3 \dfrac{\dot{\phi}^2}{H} ) ] 
\\ \nonumber
\dfrac{1}{2} (\dfrac{\nabla}{a} \delta \phi)^2 [-F^2 + \beta_0 +1 -\dfrac{\epsilon H^2}{\beta_2} (F-1+\beta_1)^2 ] \, .
\biggl{]}
\ea
As a result, the kinetic term for $\delta \phi$ vanishes in the action, yielding the decoupling of inflaton field perturbations. This is also consistent with the result in the text that $\delta \phi=B=0$. Note that the difference in the form of $\beta_2$ in Eq. (\ref{beta12}) and (\ref{beta2}) does not lead to any inconsistency, since $\beta_2$ only appears with the field $B$ in the form of $\beta_2 B$ which vanishes and does not affect the rest of equations.


\end{document}